\documentclass[envcountsame,11pt]{llncs}
\usepackage[pdftex]{graphicx}
\usepackage{epstopdf}
\usepackage{latexsym}
\usepackage{amsfonts}
\usepackage{amsmath}
\usepackage{amssymb}
\usepackage{stmaryrd}
\usepackage{color}
\usepackage[OT1]{fontenc}
\usepackage{ae,aecompl,aeguill}
\usepackage{mathrsfs, pifont}
\usepackage{microtype}
\usepackage{fixltx2e,mparhack}
\usepackage{upgreek}
\usepackage{soul}
\sethlcolor{red}

\DeclareFontFamily{OT1}{pzc}{}
\DeclareFontShape{OT1}{pzc}{m}{it}%
             {<-> s * [1.150] pzcmi7t}{}
\DeclareMathAlphabet{\mathpzc}{OT1}{pzc}%
                                 {m}{it}

\allowdisplaybreaks
\def\ut{\upharpoonright}

\def\tn#1{\textnormal{#1}}

\def\K{\tn{K}}

\def\MLR{\tn{\bf MLR}}
\def\KLR{\tn{\bf KLR}}

\def\C{\tn{C}}

\newcommand{\ord}[1]{\ensuremath{\textnormal{O}(#1)}} 

\newcommand{\emptyword}{\epsilon}

\newcommand{\TMR}{\ensuremath{\mathbf{TMR}}}
\newcommand{\TPR}{\ensuremath{\mathbf{TPR}}}
\newcommand{\TIR}{\ensuremath{\mathbf{TIR}}}
\newcommand{\PMR}{\ensuremath{\mathbf{PMR}}}
\newcommand{\PPR}{\ensuremath{\mathbf{PPR}}}
\newcommand{\PIR}{\ensuremath{\mathbf{PIR}}}

\newcommand{\Cc}[2]{\ensuremath{\mathrm{C}}\left({#1} \, \mid \, {#2}\right)}

\newcommand{\Succ}{\ensuremath{\mathrm{Succ}}}
\newcommand{\Av}[1]{\mathrm{Av}_{#1}}

\newcommand{\fs}{\ensuremath{2^{< \omega}}}
\newcommand{\prefx}{\sqsubseteq}
\newcommand{\cs}{\ensuremath{2^{\omega}}}
\newcommand{\N}{{{\mathbb{N}}}}
\newcommand{\Q}{{{\mathbb{Q}}}}

\newcommand{\uh}[1]{\upharpoonright {#1}}

\newcommand{\seqa}{A}

\newcommand{\emphdef}[1]{\textbf{#1}}

\begin{document}

\title{Separations of non-monotonic randomness notions\\
{\small (Preliminary version, 7 July 2009)}}
\author{Laurent Bienvenu \and
Rupert H\"{o}lzl
\and Thorsten Kr\"aling
\and Wolfgang Merkle
}
\institute{Institut f\"{u}r Informatik, Ruprecht-Karls-Universit\"at,\\Heidelberg, Germany}
\maketitle

\begin{abstract}
In the theory of algorithmic randomness, several notions of random sequence are defined via a game-theoretic approach, and the notions that received most attention are perhaps Martin-L\"of randomness
and computable randomness. The latter notion was introduced by Schnorr and is rather natural: an infinite binary sequence is computably random if no total computable strategy succeeds on it by betting on bits in order. However, computably random sequences can have properties that one may consider to be incompatible with being random, in particular, there are computably random sequences that are highly compressible. The concept of Martin-L\"of randomness is much better behaved in this and other respects, on the other hand its definition in terms of martingales is considerably less natural.

Muchnik, elaborating on ideas of Kolmogorov and Loveland, refined Schnorr's model by also allowing non-monotonic strategies, i.e.\ strategies that do not bet on bits in order. The subsequent ``non-monotonic'' notion of randomness, now called Kolmogorov-Loveland-randomness, has been shown to be quite close to Martin-L\"of randomness, but whether these two classes coincide remains a fundamental open question. 

In order to get a better understanding of non-monotonic randomness notions, Miller and Nies introduced some interesting intermediate concepts, where one only allows non-adaptive strategies, i.e., strategies that can still bet non-monotonically, but such that the sequence of betting positions is known in advance (and computable). Recently, these notions were shown by Kastermans and Lempp to differ from Martin-L\"of randomness. We continue the study of the non-monotonic randomness notions introduced by Miller and Nies and obtain results about the Kolmogorov complexities of initial segments that may and may not occur for such sequences, where these results then imply a complete classification of these randomness notions by order of strength.
\end{abstract}

\section{Introduction}

Random sequences are the central object of study in algorithmic randomness and have been 
investigated intensively over the last decade, which led to a wealth of interesting results clarifying the relations between the various notions of randomness and revealing interesting interactions with notions such as computational power~\cite{DowneyHirschfeldt-ta,LiVitanyi2008,Nies2009}. 

Intuitively speaking, a binary sequence is random if the bits of the sequence do not have effectively detectable regularities. This idea can be formalized in terms of betting strategies, that is, a sequence will be called random in case the capital gained by successive bets on the bits of the sequence according to a fixed betting strategy must remain bounded, with fair payoff and a fixed set of admissible betting strategies understood. 

The notions of random sequences that have received most attention are Martin-L\"of randomness and computable randomness. Here a sequence is called computably random if no total computable betting strategy can achieve  unbounded capital by betting on the bits of the sequence in the natural order, 
a definition that indeed is natural and suggests itself. However, computably random sequences may lack certain properties associated with the intuitive understanding of randomness, for example there are such sequences that are highly compressible, i.e., show a large amount of redundancy, see Theorem~\ref{thm:facile-tmr} below. Martin-L\"{o}f randomness behaves much better in this and other respects.
Indeed, the Martin-L\"{o}f random sequences can be characterized as the sequences that are incompressible in the sense that all their initial segments have essentially maximal Kolmogorov complexity, and in fact this holds for several versions of Kolmogorov complexity according to celebrated results by Schnorr, by Levin and, recently, by Miller and Yu~\cite{DowneyHirschfeldt-ta}. On the other hand, it has been held against the concept of Martin-L\"{o}f randomness that its definition involves effective approximations, i.e., a very powerful, hence rather unnatural model of computation, and indeed the usual definition of  Martin-L\"{o}f randomness in terms of left-computable martingales, that is, in terms of betting strategies where the gained capital can be effectively approximated from below, is not very intuitive.

It can be shown that Martin-L\"{o}f randomness strictly implies computable randomness. According to the preceding discussion the latter notion is too inclusive while the former may be considered unnatural. Ideally, we would therefore like to find a more natural characterization of ML-randomness; or, if that is impossible, we are alternatively interested in a notion that is close in strength to ML-randomness, but has a more natural definition. One promising way of achieving such a more natural characterization or definition could be to use computable betting strategies that are more powerful than those used to define computable randomness.

Muchnik~\cite{MuchnikSU1998} proposed to consider computable betting strategies that are non-monotonic in the sense that the bets on the bits need not be done in the natural order, but such that the bit to bet on next can be computed from the already scanned bits. The corresponding notion of randomness is called Kolmogorov-Loveland randomness because 
Kolmogorov and Loveland independently had proposed concepts of randomness defined via non-monotonic selecting of bits. 

Kolmogorov-Loveland randomness is implied by and in fact is quite close to Martin-L\"of randomness, see Theorem~\ref{thm:muchnik-ppr} below, but whether the two notions are distinct is one of the major open problems of algorithmic randomness. In order to get a better understanding of this open problem and of non-monotonic randomness in general, Miller and Nies~\cite{MillerN2006} introduced 
restricted variants of Kolmogorov-Loveland randomness, where the sequence of betting positions must be non-adaptive, i.e., can be computed in advance without knowing the sequence on which one bets.

The randomness notions mentioned so far are determined by two parameters that correspond to the columns and rows, respectively, of the table in Figure~\ref{figure:knowninclusions}.
First, the sequence of places that are scanned and on which bets may be placed, while always being given effectively, can just be monotonic, can be equal to~$\pi(0), \pi(1), \ldots$ for a permutation or an injection~$\pi$ from~$\N$ to~$\N$, or can be  adaptive, i.e., the next bit depends on the bits already scanned. Second, once the sequence of scanned bits is determined, betting on these bits can be according to a betting strategy where the corresponding martingale is total or partial computable, or is left-computable. The known inclusions between the corresponding classes of random sequences are shown in Figure~\ref{figure:knowninclusions}, see Section~\ref{sec:permutation} for technical details and for the definitions of the class acronyms that occur in the figure. 
\setlength{\fboxsep}{2ex}
\begin{figure}[h]
\begin{center}
\framebox{%
\begin{tabular}{r|ccccccc}
  & monotonic & & permutation & & injection & & adaptive\\
 \hline

 total\, & \TMR & $=$ & \TPR & $\supseteq$ &  \TIR & $\supseteq$ & \KLR\\
          &  \rotatebox{90}{$\subseteq$}& & \rotatebox{90}{$\subseteq$} & & \rotatebox{90}{$\subseteq$}& & \rotatebox{90}{$=$}\\
 partial\, & \PMR &$\supseteq$ & \PPR & $\supseteq$ & \PIR & $\supseteq$ &\KLR\\
 
           &  \rotatebox{90}{$\subseteq$}& & \rotatebox{90}{$\subseteq$} & & \rotatebox{90}{$\subseteq$} &&\rotatebox{90}{$\subseteq$}\\

   left-computable\, & \MLR & = & \MLR & = & \MLR & = &\MLR\\ 
\end{tabular}
\medskip
}
\end{center}
 \caption{Known class inclusions \label{figure:knowninclusions}}
\end{figure}

The classes in the last row of the table in Figure~\ref{figure:knowninclusions} all  coincide with the class of Martin-L\"{o}f  random sequences by the folklore result that left-computable martingales always yield the concept of Martin-L\"{o}f randomness, no matter whether the sequence of bits to bet on is monotonic or is determined adaptively, because even in the latter, more powerful model one can uniformly in~$k$ enumerate an open cover of measure at most~$1/k$ for all the sequences  on which some universal martingale exceeds~$k$. Furthermore, the classes in the first and second row of the last column coincide with the class  of Kolmogorov-Loveland random sequences, because it can be shown that 
total and partial adaptive betting strategies yield the same concept of random sequence~\cite{Merkle2003}. Finally, it follows easily from results of Buhrman et al.~\cite{BuhrmanvMRSS2000} that the class~\TMR\ of computably random sequences coincides with the class~\TPR\ of sequences that are random with respect to total permutation martingales, i.e., the ability to scan the bits of a sequence according to a computable permutation does not increase the power of total martingales. 

Concerning non-inclusions, it is well-known that it holds that \[\KLR \subsetneq \PMR \subsetneq \TMR.\]
Furthermore, Kastermans and Lempp~\cite{KastermansL} have recently shown that the Martin-L\"{o}f random sequences form a proper subclass of the class~\PIR\ of partial injective random sequences, i.e., $\MLR \subsetneq\PIR$.

Apart from trivial consequences of the definitions and the results just mentioned, nothing has been known about the relations of the randomness notions between computable randomness and 
Martin-L\"{o}f randomness in Figure~\ref{figure:knowninclusions}.
In what follows, we investigate the six randomness notions that are shown in Figure~\ref{figure:knowninclusions} in the range between~\PIR\ and~\TMR, i.e., between partial injective randomness as introduced below and computable randomness.  
We obtain a complete picture of the inclusion structure of these notions, more precisely we show that the notions are mutually distinct and indeed are mutually incomparable with respect to set theoretical inclusion, except for the inclusion relations that follow trivially by definition and by the known relation~$\TMR\subseteq \TPR$, see Figure~\ref{figure:inclusionstructure} at the end of this paper. Interestingly these separation results are obtained by investigating the possible values of the Kolomogorov complexity of initial segments of random sequences for the different strategy types, and for some randomness notions we obtain essentially sharp bounds on how low these complexities can be. 

\paragraph{Notation.}
We conclude the introduction by fixing some notation. The set of finite strings (or finite binary sequences, or words) is denoted by $\fs$, $\emptyword$ being the empty word. We denote the set of infinite binary sequences by $\cs$. Given two finite strings~$w,w'$, we write $w \prefx w'$ if~$w$ is a prefix of~$w'$. Given an element $x$ of $\cs$ or $\fs$, $x(i)$ denotes the $i$-th bit of~$x$ (where by convention there is a $0$-th bit and $x(i)$ is undefined if~$x$ is a word of length less than $i+1$). If $\seqa \in \cs$ and $X=\{x_0<x_1<x_2<\ldots\}$ is a subset of~$\N$ then $\seqa \uh{X}$ is the finite or infinite binary sequence $\seqa(x_0)\seqa(x_1)\ldots$. We abbreviate $\seqa \uh{\{0, \ldots, n-1\}}$ by $\seqa \uh{n}$ (i.e., the prefix of $\seqa$ of length~$n$). 

$\C$ and $\K$ denote plain and prefix-free Kolmogorov complexity, respectively~\cite{DowneyHirschfeldt-ta,LiVitanyi2008}. The function $\log$ designates the logarithm of base 2. An \emphdef{order} is a function $h: \N \rightarrow \N$ that is non-decreasing and tends to infinity.
\section{Permutation and injection randomness}\label{sec:permutation}
We now review the concept of martingale and betting strategy that are central for the unpredictability approach to define notions of an infinite random sequence.  

\begin{definition}\label{def:martingale}
A \emphdef{martingale} is a nonnegative, possibly partial, function $d: \fs \rightarrow \Q$ such that for all $w \in \fs$, if $d(w0)$ is defined if and only if $d(w1)$ is, and if these are defined, then so is $d(w)$, and the relation $2d(w)=d(w0)+d(w1)$ holds. A martingale \emphdef{succeeds} on a sequence $\seqa \in \cs$ if $d(\seqa \uh{n})$ is defined for all~$n$, and $\limsup\, d(\seqa \uh{n}) = + \infty$.  We denote by $\Succ(d)$ the \emphdef{success set} of~$d$, i.e., the set of sequences on which~$d$ succeeds. 
\end{definition}

Intuitively, a martingale represents the capital of a player who bets on the bits of a sequence $\seqa \in \cs$ in order, where at every round she bets some amount of money on the value of the next bit of $\seqa$. If her guess is correct, she doubles her stake. If not, she loses her stake. The quantity~$d(w)$, with~$w$ a string of length~$n$, represents the capital of the player before the~$n$-th round of the game (by convention there is a $0$-th round) when the first~$n$ bits revealed so far are those of~$w$.\\

We say that a sequence $\seqa$ is \emphdef{computably random} if no total computable martingale succeeds on it. One can extend this in a natural way to partial computable martingales: a sequence~$\seqa$ is \emphdef{partial computably random} if no partial martingale succeeds on it. No matter whether we consider partial or total computable martingales, this game model can be seen as too restrictive by the discussion in the introduction. Indeed, one could allow the player to bet on bits in any order she likes (as long as she can visit each bit at most once). This leads us to extend the notion of martingale to the notion of strategy.  

\begin{definition}
A \emphdef{betting strategy} is a pair $b=(d,\sigma)$ where $d$ is a martingale and $\sigma: \fs \rightarrow \N$ is a function. 
\end{definition}

For a strategy $b=(d,\sigma)$, the term~$\sigma$ is called the \emph{scan rule}. For a string~$w$, $\sigma(w)$ represents the position of the next bit to be visited if the player has read the sequence of bits~$w$ during the previous moves. And as before, $d$ specifies how much money is bet at each move. Formally, given an~$\seqa \in \cs$, we define by induction a sequence of positions $n_0,n_1,\ldots$ by
$$
\left\{ \begin{array}{l} n_0=\sigma(\emptyword), \\ n_{k+1}=\sigma \left(A(n_0)A(n_1)\ldots A(n_k)\right) \tn{ for all } k \geq 0\end{array} \right.
$$ 
and we say that $b=(d,\sigma)$ \emphdef{succeeds} on $A$ if the $n_i$ are all defined and pairwise distinct (i.e., no bit is visited twice) and
$$
\limsup_{k \rightarrow +\infty}\; d\left(A(n_0)\ldots A(n_k)\right)=+\infty
$$

Here again, a betting strategy $b=(d,\sigma)$ can be total or partial. In fact, its partiality can be due either to the partiality of~$d$ or to the partiality of~$\sigma$. We say that a sequence is \emphdef{Kolmogorov-Loveland random} if no total computable betting strategy succeeds on it. As noted in~\cite{MerkleMNRS2006}, the concept of Kolmogorov-Loveland randomness remains the same if one replaces ``total computable'' by ``partial computable'' in the definition.

Kolmogorov-Loveland randomness is implied by Martin-L\"of randomness and whether the two notions can be separated is one of the most important open problems on algorithmic randomness. As we discussed above, Miller and Nies~\cite{MillerN2006} proposed to look at intermediate notions of randomness, where the power of non-monotonic betting strategies is limited. In the definition of a betting strategy, the scan rule is adaptive, i.e., the position of the next visited bit depends on the bits previously seen. It is interesting to look at non-adaptive games. 

\begin{definition}\label{def:injectionstrategy}
In the above definition of a strategy, when $\sigma(w)$ only depends on the length of~$w$ for all $w$ (i.e., the decision of which bit should be chosen at each move is independent of the values of the bits seen in previous moves), we identify~$\sigma$ with the (injective) function~$\pi:Â \N \rightarrow \N$, where for all~$n$ $\pi(n)$ is the value of $\sigma$ on words of length~$n$ ($\pi(n)$ indicates the position of the bit visited during the $n$-th move), and we say that $b=(d,\pi)$ is an \emphdef{injection strategy}. If moreover~$\pi$ is bijective, we say that~$b$ is a \emphdef{permutation strategy}. If $\pi$ is the identity, the strategy $b=(d,\pi)$ is said to be \emphdef{monotonic}, and can clearly be identified with the martingale~$d$. 
\end{definition}

All this gives a number of possible non-adaptive, non-monotonic, randomness notions: one can consider either monotonic, permutation, or injection strategies, and either total computable or partial computable ones. This gives a total of six randomness classes, which we denote by 
\begin{equation}\label{eq:classes}
\TMR,\; \TPR, \;\TIR,\; \PMR, \;\PPR, \;\mbox{and} \;\PIR , 
\end{equation}
where the first letter indicates whether we consider total (T) or partial (P) strategies, and the second indicates whether we look at monotonic (M), permutation (P) or injection (I) strategies. For example, the class $\TMR$ is the class of computably random sequences, while the class $\PIR$ is the class of sequences~$\seqa$ such that no partial injection strategy succeeds on~$\seqa$. 
Recall in this connection that the known inclusions between the six classes in~\eqref{eq:classes} and the classes~\KLR\  and~\MLR\ of Kolmogorov-Loveland random and Martin-L\"{o}f random sequences have been shown in Figure~\ref{figure:knowninclusions} above. 

\section{Randomness notions based on total computable strategies}

We begin our study by the randomness notions arising from the game model where strategies are total computable. As we will see, in this model, it is possible to construct sequences that are random and yet have very low Kolmogorov complexity (i.e.\ all their initial segments are of low Kolmogorov complexity). We will see in the next section that this is no longer the case when we allow partial computable strategies in the model. 

\subsection{Building a sequence in $\TMR$ of low complexity}

The following theorem is a first illustration of the phenomenon we just described. 

\begin{theorem}[Lathrop and Lutz~\cite{LathropL1999}, Muchnik \cite{MuchnikSU1998}] 
 \label{thm:facile-tmr} For every computable order $h$, there is a sequence $\seqa \in \mathbf{\TMR}$ such that, for all $n \in \mathbb{N}$, $$\Cc{A\uh{n}}{n} \leq h(n)+ \ord{1}. $$
\end{theorem} 

\begin{proof}[Idea]
Defeating one total computable martingale is easy and can be done computably, i.e., for every total computable martingale~$d$ there exists a sequence~$\seqa$, uniformly computable in~$d$, such that $\seqa \notin \Succ(d)$. Indeed, given a martingale~$d$. For any given~$w$, one has either~$d(w0) \leq d(w)$ or $d(w1) \leq d(w)$. Thus, one can easily construct a computable sequence~$\seqa$ by setting~$\seqa \uh{0}= \emptyword$ and by induction, having defined $\seqa \uh{n}$, we choose $\seqa \uh{n+1}=(\seqa \uh{n})i$ where $i \in \{0,1\}$ is such that~$d((\seqa \uh{n})i) \leq d(\seqa \uh{n})$. This can of course be done computably since~$d$ is total computable, and by construction of~$\seqa$, $d(\seqa \uh{n})$ is non-increasing, meaning in particular that~$d$ does not succeed against~$\seqa$.\\

Defeating a finite number of total computable martingales is equally easy. Indeed, given a finite number~$d_1,\ldots,d_k$ of such martingales, their sum~$D=d_1+\ldots+d_k$ is itself a total computable martingale (this follows directly from the definition). Thus, we can construct as above a computable sequence~$A$ that defeats~$D$. And since $D \geq d_i$ for all $1 \leq i \leq k$, this implies that $A$ defeats all the~$d_i$. Note that this argument would work just as well if we had taken~$D$ to be any weighted sum $\alpha_1 d_1 + \ldots + \alpha_k d_k$, with positive rational constants~$\alpha_i$.\\ 

We now need to deal with the general case where we have to defeat \emph{all} total computable martingales simultaneously. We will again proceed using a diagonalization technique. Of course, this diagonalization cannot be carried out effectively, since there are infinitely many such martingales and since we do not even know whether any one given partial computable martingale is total. The first problem can easily be overcome by introducing the martingales to diagonalize against one by one instead of all at the beginning. So at first, for a number of stages we will only take into account the first computable martingale~$d_1$. Then (maybe after a long time) we may introduce the second martingale~$d_2$, with a small coefficient $\alpha_2$ (to ensure that introducing~$d_2$ does not cost us too much) and then consider the martingale~$d_1+\alpha_2 d_2$. Much later we can introduce the third martingale~$d_3$ with an even smaller coefficient~$\alpha_3$, and diagonalize against $d_1+\alpha_2 d_2 + \alpha_3 d_3$, and so on. So in each step of the construction we have to consider just a finite number of martingales.

The non-effectivity of the construction arises from the second problem, deciding which of our partial computable martingales are total. However, once we are supplied with this additional information, we {\it can} effectively carry out the construction of~$\seqa$. And since for each step we need to consider only finitely many potentially total martingales, the information we need to construct the first~$n$ bits of~$\seqa$ for some fixed~$n$ is finite, too. Say, for example, that for the first~$n$ stages of the construction -- i.e., to define $\seqa \uh{n}$ -- we decided on only considering~$k$ martingales $d_0,\ldots,d_k$. Then we need no more than~$k$ bits, carrying the information which martingales among $d_0,\ldots,d_k$ are total, to describe $\seqa \uh{n}$. That way, we get $\Cc{\seqa \uh{n}}{n} \leq k+O(1)$.

As can be seen from the above example, the complexity of descriptions of prefixes of $A$ depends on how fast we introduce the martingales. 
This is where our orders come into play. Fix a fast-growing computable function~$f$ with $f(0)=0$, to be specified later. We will introduce a new martingale at every position of type $f(k)$, that is, between positions~$[f(k),f(k+1))$, we will only diagonalize against~$k+1$ martingales, hence by the above discussion, for every~$n \in [f(k),f(k+1))$, we have
$$
\Cc{\seqa \uh{n}}{n} \leq k+O(1)
$$
Thus, if the function~$f$ grows faster than the inverse function~$h^{-1}$ of a given order~$h$, we get
$$
\Cc{\seqa \uh{n}}{n} \leq h(n)+O(1)
$$
 for all~$n$.\qed
\end{proof}

\subsection{$\TMR=\TPR$: the averaging technique}

It turns out that, perhaps surprisingly, the classes $\TMR$ and $\TPR$ coincide. This fact was stated explicitely in Merkle et al~\cite{MerkleMNRS2006}, but is easily derived from the ideas introduced in Buhrman et al~\cite{BuhrmanvMRSS2000}. We present the main ideas of their proof as we will later need them. We shall prove:

\begin{theorem}\label{thm:permutation-monotonization}
Let~$b=(d,\pi)$ be a total computable permutation strategy. There exists a total computable martingale~$d$ such that $\Succ(b) \subseteq \Succ(d)$.
\end{theorem}

This theorem states that total permutation strategies are no more powerful than total monotonic strategies, which obviously entails~$\TMR=\TPR$. Before we can prove it, we first need a definition.

\begin{definition}
Let $b=(d,\pi)$ be a total injective strategy. Let $w \in \fs$. We can run the strategy~$b$ on~$w$ as if it were an element of $\cs$, stopping the game when $b$ asks to bet on a bit of position outside~$w$. This game is of course finite (for a given~$w$) since at most~$|w|$ bets can be made. We define~$\hat{b}(w)$ to be the capital of~$b$ at the end of this game. Formally: $\hat{b}(w)=d\left(w_{\pi(0)} \ldots w_{\pi(N-1)}\right)$ where~$N$ is the smallest integer such that $\pi(N) \geq |w|$. 
\end{definition}

Note that if $b=(d,\pi)$ is a total computable injection martingale, $\hat{b}$ is total computable. If $\hat b$ was itself a monotonic martingale, Theorem~\ref{thm:permutation-monotonization} would be proven. This is however not the case in general: suppose $d(\emptyword)=1$, $d(0)=2$, $d(1)=0$, and $\pi(0)=1$, $\pi(1)=5$ (i.e., $d$ first visits the bit in position~$1$, betting everyrhing on the value~$0$, then visits the bit in position~$5$). We then have~$b(0)=1$ and $b(1)=1$, but $\hat{b}(00)=2$, $\hat{b}(01)=2$, $\hat{b}(10)=0$ and $\hat{b}(11)=0$, which shows that~$\hat{b}$ is not a martingale.\\

The trick is, given a betting strategy~$b$ and a word~$w$, to look at the \emph{expected value} of~$b$ on~$w$, i.e., look at the mathematical expectation of~$b(w')$ for large enough extensions~$w'$ of~$w$. Specifically, given a total betting strategy $b=(d,\pi)$ and a word~$w$ of length~$n$, we take an integer~$M$ large enough to have $$\pi\left([0,\ldots,M-1]\right) \cap [0,\ldots,n-1] = \pi(\N) \cap  [0,\ldots,n-1]$$ (i.e.\ the strategy~$b$ will never bet on a bit of position less than~$n$ after the $M$-th move), and define:
$$
\Av{b}(w)=\frac{1}{2^M}\, \sum_{\substack{w \prefx w'\\|w'|=M}} \hat{b}(w')
$$

\begin{proposition}[Buhrman et al~\cite{BuhrmanvMRSS2000}, Kastermans-Lempp~\cite{KastermansL}]\label{prop:avg}
\begin{itemize}
\item[(i)] The quantity $\Av{b}(w)$ (defined above) is well-defined i.e.\ does not depend on~$M$ as long as it satisfies the required condition.
\item[(ii)] For a total injective strategy~$b$, $\Av{b}$ is a martingale.
\item[(iii)] For a given injective strategy~$b$ and a given word~$w$ of length~$n$, $\Av{b}(w)$ can be computed if we know the set $\pi(\N) \cap [0,\ldots,n-1]$. In particular, if~$b$ is a total computable permutation strategy, then $\Av{b}$ is total computable. 
\end{itemize}
\end{proposition}

As Buhrman et al.~\cite{BuhrmanvMRSS2000} explained, it is not true in general that if a total computable injective strategy~$b$ succeeds against a sequence~$\seqa$, then~$\Av{b}$ also succeeds on~$\seqa$. However, this can be dealt with using the well-known ``saving trick''. Suppose we are given a martingale~$d$ with initial capital, say, $1$. Consider the variant~$d'$ of $d$ that does the following: when run on a given sequence~$\seqa$, $d'$ initially plays exactly as~$d$. If at some stage of the game~$d'$ reaches a capital of~$2$ or more, it then puts half of its capital on a ``bank account'', which will never be used again. From that point on, $d'$ bets half of what~$d$ does, i.e.\ start behaving like $d/2$ (plus the saved capital). If later in the game the ``non-saved'' part of its capital reaches~$2$ or more, then half of it is placed on the bank account and then~$d'$ starts behaving like~$d/4$, and so on.\\

\label{page:saving} For every martingale~$d'$ that behaves as above (i.e.\ saves half of its capital as soon as it exceeds twice its starting capital), we say that~$d'$ has the ``saving property''. It is clear from the definition that if~$d$ is computable, then so is~$d'$, and moreover~$d'$ can be uniformly computed given an index for~$d$. Moreover, if for some sequence~$\seqa$ one has
$$
\limsup_{n \rightarrow +\infty} d(\seqa \uh{n}) = +\infty
$$
then
$$
\lim_{n \rightarrow +\infty} d'(\seqa \uh{n}) = +\infty
$$
which in particular implies $\Succ(d) \subseteq \Succ(d')$ (it is easy to see that it is in fact an equality). Thus, whenever one considers a martingale~$d$, one can assume without loss of generality that it has the saving property (as long as we are only interested in the success set of martingales, not in the growth rate of their capital). The key property (for our purposes) of saving martingales is the following.

\begin{lemma}\label{lem:saving-avg}
Let~$b=(d,\pi)$ be a total injective strategy such that~$d$ has the saving property. Let $d'=\Av{b}$. Then $\Succ(b) \subseteq \Succ(d')$. 
\end{lemma}

\begin{proof}
Suppose that $b=(d,\pi)$ succeeds on a sequence~$\seqa$. Since~$d$ has the saving property, for arbitrarily large $k$ there exists a finite prefix $\seqa \uh{n}$ of~$A$ such that a capital of at least~$k$ is saved during the finite game of~$b$ against~$\seqa$. We then have $\hat{b}(w') \geq k$ for all extensions $w'$ of $\seqa \uh{n}$ (as a saved capital is never used), which by definition of $\Av{b}$ implies $\Av{b}(\seqa \uh{m}) \geq k$ for all~$m \geq n$. Since~$k$ can be chosen arbitrarily large, this finishes the proof. \qed
\end{proof}

Now the proof of Theorem~\ref{thm:permutation-monotonization} is as follows. Let~$b=(d,\pi)$ be a total computable permutation strategy. By the above discussion, let~$d'$ be the saving version of~$d$, so that $\Succ(d) \subseteq \Succ(d')$. Setting $b'=(d',\pi)$, we have $\Succ(b) \subseteq \Succ(b')$. By Proposition~\ref{prop:avg} and Lemma~\ref{lem:saving-avg}, $d''=\Av{b'}$ is a total computable martingale, and $$\Succ(b) \subseteq \Succ(b') \subseteq \Succ(d'')$$ as wanted.

\subsection{Understanding the strength of injective strategies: the class $\TIR$}

While the class of computably random sequence (i.e.\ the class $\TMR$) is closed under computable permutations of the bits, we now see that this result does not extend to computable injections. To wit, the following theorem is true. 

\begin{theorem}\label{thm:tir1}
Let $\seqa \in \cs$. Let $\{n_k\}_{k \in \N}$ be a computable sequence of integers such that $n_{k+1}\geq 2 n_k$ for all~$k$. 
Suppose that $\seqa$ is such that:
$$
\Cc{\seqa \uh{n_k}}{k} \leq \log(n_k)-3\log(\log(n_k))
$$
for infinitely many~$k$. Then $\seqa \notin \TIR$.
\end{theorem}

\begin{proof}
Let $\seqa$ be a sequence satisfying the hypothesis of the theorem. Assuming, without loss of generality, that $n_0=0$, we partition $\N$ into an increasing sequence of intervals $I_0,I_1,I_2,\ldots$ where $I_k=[n_k,n_{k+1})$.
Notice that we have for all~$k$:
$$
\Cc{\seqa \uh I_k}{k} \leq \Cc{\seqa \uh n_{k+1}}{k+1}+O(1)
$$

By the hypothesis of the theorem, the right-hand side of the above inequality is bounded by $\log(n_{k+1})-3\log(\log(n_{k+1}))$ for infinitely many~$k$.\\

 Additionally, we have $|I_k|=n_{k+1}-n_k$ which by hypothesis on the sequence $n_k$ implies $|I_k| \geq n_{k+1}/2$, and hence $\log(|I_k|)=\log(n_{k+1})+O(1)$ and  $\log(\log(|I_k|))=\log(\log(n_{k+1}))+O(1)$. It follows that
$$
\Cc{\seqa \uh I_k}{k} \leq  \log(|I_k|)-3\log(\log(|I_k|))-O(1)
$$
for infinitely many~$k$, hence
$$
\Cc{\seqa \uh I_k}{k} \leq  \log(|I_k|)-2\log(\log(|I_k|))
$$
for infinitely many~$k$.

Let us call $S_k$ the set of strings~$w$ of length $|I_k|$ such that $\Cc{w}{|I_k|} \leq \log(|I_k|)-2\log(\log(|I_k|))$ (to which $\seqa \uh{I_k}$ belongs for infinitely many~$k$). By the standard counting argument, there are at most $$s_k=2^{\log(|I_k|)-2\log(\log(|I_k|))}=\frac{|I_k|}{\log^2(|I_k|)}$$ strings in $S_k$. For every $k$, we split $I_k$ into $s_k$ consecutive disjoint intervals of equal length:
$$
I_k=J^0_k \cup J^1_k \cup \ldots \cup J^{s_k-1}_k
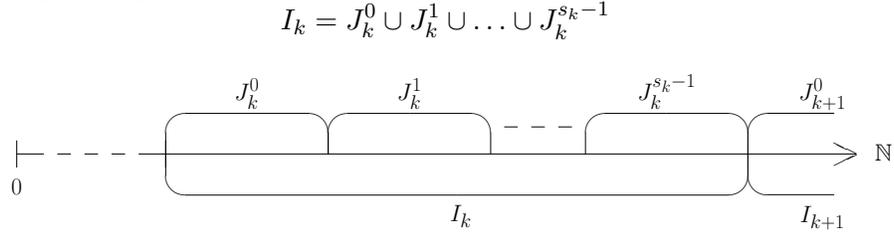
$$

\begin{center}
\scalebox{0.36}{ \setlength{\unitlength}{4144sp}%
\begingroup\makeatletter\ifx\SetFigFont\undefined
\def\x#1#2#3#4#5#6#7\relax{\def\x{#1#2#3#4#5#6}}%
\expandafter\x\fmtname xxxxxx\relax \def\y{splain}%
\ifx\x\y   
\gdef\SetFigFont#1#2#3{%
  \ifnum #1<17\tiny\else \ifnum #1<20\small\else
  \ifnum #1<24\normalsize\else \ifnum #1<29\large\else
  \ifnum #1<34\Large\else \ifnum #1<41\LARGE\else
     \huge\fi\fi\fi\fi\fi\fi
  \csname #3\endcsname}%
\else
\gdef\SetFigFont#1#2#3{\begingroup
  \count@#1\relax \ifnum 25<\count@\count@25\fi
  \def\x{\endgroup\@setsize\SetFigFont{#2pt}}%
  \expandafter\x
    \csname \romannumeral\the\count@ pt\expandafter\endcsname
    \csname @\romannumeral\the\count@ pt\endcsname
  \csname #3\endcsname}%
\fi
\fi\endgroup
\begin{picture}(14271,2805)(-11,-7679)
\put(14260,-6473){\makebox(0,0)[lb]{\smash{\SetFigFont{29}{34.8}{rm}$\mathbb{N}$}}}
\thicklines
\put(13585,-6361){\oval(2868,1354)[tl]}
\put(7313,-6361){\oval(9674,1368)[bl]}
\put(7313,-6361){\oval(9676,1368)[br]}
\put(3826,-6361){\oval(2700,1350)[tr]}
\put(3826,-6361){\oval(2700,1350)[tl]}
\put(6526,-6361){\oval(2700,1350)[tr]}
\put(6526,-6361){\oval(2700,1350)[tl]}
\put(10801,-6361){\oval(2700,1350)[tr]}
\put(10801,-6361){\oval(2700,1350)[tl]}
\put(13585,-6557){\line( 2, 1){371.200}}
\put(13951,-6361){\line(-2, 1){371.600}}
\put(2026,-6361){\line( 1, 0){11925}}
\put(  1,-6136){\line( 0,-1){450}}
\put(  1,-6361){\line( 1, 0){450}}
\put(676,-6361){\line( 1, 0){225}}
\put(1126,-6361){\line( 1, 0){225}}
\put(1576,-6361){\line( 1, 0){225}}
\put(2476,-6136){\line( 0,-1){450}}
\put(12151,-6136){\line( 0,-1){450}}
\put(8101,-5911){\line( 1, 0){225}}
\put(8551,-5911){\line( 1, 0){225}}
\put(9001,-5911){\line( 1, 0){225}}
\put(13388,-7500){\makebox(0,0)[b]{\smash{\SetFigFont{29}{34.8}{rm}$I_{k+1}$}}}
\put(13388,-5461){\makebox(0,0)[b]{\smash{\SetFigFont{29}{34.8}{rm}$J_{k+1}^0$}}}
\put(10801,-5461){\makebox(0,0)[b]{\smash{\SetFigFont{29}{34.8}{rm}$J_k^{s_k-1}$}}}
\put(6526,-5461){\makebox(0,0)[b]{\smash{\SetFigFont{29}{34.8}{rm}$J_k^1$}}}
\put(3826,-5461){\makebox(0,0)[b]{\smash{\SetFigFont{29}{34.8}{rm}$J_k^0$}}}
\put(7201,-7486){\makebox(0,0)[lb]{\smash{\SetFigFont{29}{34.8}{rm}$I_k$}}}
\put(  1,-7036){\makebox(0,0)[b]{\smash{\SetFigFont{29}{34.8}{rm}0}}}
\put(13585,-6361){\oval(2868,1354)[bl]}
\end{picture}
} 
\end{center}

We design a betting strategy as follows. We start with a capital of~$2$. We then reserve for each $k$ an amount $1/(k+1)^2$ to be bet on the bits in positions in $I_k$ (this way, the total amount we distribute is smaller than~$2$), and we split this evenly between the $J^i_k$, i.e.\ we reserve an amount $\frac{1}{ s_k \cdot (k+1)^2 }$ for every $J^i_k$. We then enumerate the sets $S_k$ in parallel. Whenever the $e$-th element $w^e_k$ of some $S_k$ is enumerated, we see $w^e_k$ as a possible candidate to be equal to $\seqa \uh{I_k}$, and we bet the reserved amount  $\frac{1}{s_k \cdot(k+1)^2}$ on the fact that $\seqa \uh{I_k}$ coincides with $w^e_k$ on the bits whose position is in $J^e_k$. If we are successful (this in particular happens whenever $w^e_k=\seqa \uh{I_k}$), our reserved capital for this $J^e_k$ is multiplied by $2^{|J^e_k|}$, i.e.\ we now have for this $J^e_k$, a capital of
$$
\frac{1}{s_k \cdot (k+1)^2} \cdot 2^{(|I_k|/s_k)}
$$
Replacing $s_k$ by its value (and remembering that $|I_k|\geq 2^{k-O(1)}$), an elementary calculation shows that this quantity is greater than~$1$ for almost all~$k$. Thus, our betting strategy succeeds on $\seqa$. Indeed, for infinitely many~$k$, $\seqa \uh{I_k}$ is an element of $S_k$, hence for some $e$ we will be successful in the above sub-strategy, making an amount of money greater than~$1$ for infinitely many~$k$, hence our capital tends to infinity throughout the game. Finally, it is easy to see that this betting strategy is total: it simply is a succession of doubling strategies on an infinite~c.e.\ set of words, and it is injective as the $J^e_k$ form a partition of $\N$, and the order of the bits we bet on is independent of $\seqa$ (in fact, we see our betting strategy succeeds on \emph{all} sequences $\alpha$ satisfying the hypothesis of the theorem). \qed

\end{proof}

As an immediate corollary, we get the following.

\begin{corollary}
 If for a sequence~$A$ we have for all~$n$ $\Cc{\seqa \uh{n}}{n} < \log n - 4 \log\log n + \ord{1} $, then $A\not\in\TIR$.
\end{corollary}

Another interesting corollary of our construction is that the class of all computable sequences can be covered by a single total computable injective strategy.

\begin{corollary}
There exists a single total computable injective strategy which succeeds against all computable elements of~$\cs$.
\end{corollary}

\begin{proof}
This is because, as we explained above, the strategy we construct in the proof of Theorem~\ref{thm:tir1} succeeds against \emph{every} sequence~$A$ such that $\Cc{A \uh {n_k}}{k} \leq \log(n_k)-3\log(\log(n_k))$ for infinitely many~$k$. This in particular includes all computable sequences~$A$, for which $\Cc{A \uh {n_k}}{k}=O(1)$. \qed
\end{proof}

The lower bound on Kolmogorov complexity given in Theorem~\ref{thm:tir1} is quite tight, as witnessed by the following theorem.

\begin{theorem}\label{thm:tir2}
For every computable order $h$ there is a sequence $A\in \TIR$ such that $\C(A\ut n\mid n)\leq \log(n) + h(n) + \ord{1}$. In particular, we have $\C(A\uh{n}) \leq 2\log(n) + h(n) + \ord{1}$.
\end{theorem}


\begin{proof}
The proof is a modification of the proof of Theorem~\ref{thm:facile-tmr}. This time, we want to diagonalize against all \emph{non-monotonic} total computable injective betting strategies. Like in the proof of Theorem~\ref{thm:facile-tmr}, we add them one by one, discarding the partial strategies. However, to achieve the construction of~$\seqa$ by diagonalization, we will diagonalize against the average martingales of the strategies we consider. As explained on page~\pageref{page:saving}, we can assume that all total computable injective strategies have the saving property, hence defeating $\Av{b}$ is enough to defeat~$b$ (by Lemma~\ref{lem:saving-avg}). The proof thus goes as follows: 

Fix a fast growing computable function~$f$, to be specified later. We start with a martingale~$D_0=1$ (the constant martingale equal to~$1$) and $w_0=\emptyword$. For all~$k$ we do the following. Assume we have constructed a prefix $w_k$ of~$\seqa$ of length $f(k)$, and that we are currently diagonalizing against a martingale~$D_k$, so that $D_k(w_k)<2$. We then enumerate a new partial computable injective betting strategy~$b$. If it is not total, we memorize this fact using one extra bit of information, and we set $D_{k+1}=D_k$. Otherwise, we set $d_{k+1}=\Av{b}$ and compute a positive rational $\alpha_{k+1}$ such that $(D_k+\alpha_{k+1}d_{k+1})(w_k) < 2$, and finally set $D_{k+1}=D_k+\alpha_{k+1}d_{k+1}$.

Then, we define~$w_{k+1}$ to be the extension of $w_k$ of length~$f(k+1)$ by the usual diagonalization against~$D_{k+1}$, maintaining the inequality $D_{k+1}(u) < 2$ for all prefixes~$u$ of~$w_{k+1}$. The infinite sequence~$\seqa$ obtained this way defeats all the average martingales of all total computable injective strategies, hence by Lemma~\ref{lem:saving-avg}, $\seqa \in \TIR$.\\

It remains to show that~$\seqa$ has low Kolmogorov complexity. Suppose we want to describe~$\seqa \uh{n}$ for some~$n \in [f(k),f(k+1))$. This can be done by giving~$n$, the subset of $\{0,\ldots,k\}$ (of complexity $k+O(1)$) corresponding to the indices of the total computable injective strategies among the first~$k$ partial computable ones, and by giving the restriction of $D_{k+1}$ to words of length at most~$n$. From all this, $\seqa \uh{n}$ can be reconstructed following the above construction. It remains to evaluate the complexity of the restriction of $D_{k+1}$ to words of length at most~$n$. We already know the total computable injective strategies $b_0,\ldots,b_{k}$ that are being considered in the definition of~$D_{k+1}$. For all~$i$, let $\pi_i$ be the injection associated to $b_i$. We need to compute, for all~$0 \leq i \leq k$, the martingale~$d_i=\Av{b_i}$ on words of length at most~$n$. By Proposition~\ref{prop:avg}, this can be done knowing $\pi_i(\N) \cap [0,\ldots,n-1]$ for all~$0 \leq i \leq k$. But if the $\pi_i$ are known, this set is uniformly c.e.\ in $i,n$. Hence, we can enumerate all the sets  $\pi_i(\N) \cap [0,\ldots,n-1]$ (for~$0 \leq i \leq k$) in parallel, and simply give the last couple $(i,l)$ such that $l$ is enumerated in $\pi_i(\N) \cap [0,\ldots,n-1]$. Since $0 \leq i \leq k$ and $0 \leq l < n$, this costs an amount of information $\ord{\log k} + \log n$. To sum up, we get
$$
\Cc{\seqa \uh{n}}{n} \leq k + \ord{\log k} + \log n
$$ 
Thus, it suffices to take~$f$ growing fast enough to ensure that the term $\leq k + \ord{\log k}$ is smaller than $h(n)+\ord{1}$. \qed

%


\end{proof}

\section{Randomness notions based on partial computable strategies}

We now turn our attention to the second line of Figure~\ref{figure:knowninclusions}, i.e., to those randomness notions that are based on partial computable betting strategies.

\subsection{The class $\PMR$: partial computable martingales are stronger than total ones}

We have seen in the previous section that some sequences in $\TIR$ (and a fortiori $\TPR$ and $\TMR$) may be of very low complexity, namely logarithmic. This is not the case anymore when one allows partial computable strategies, even monotonic ones. 

\begin{theorem}[Merkle~\cite{Merkle2008}]\label{theorem-merkle}
If $\C(A\uh{n}) = \ord{\log n}$ then $\seqa \not\in \PMR$.
\end{theorem}

However, the next theorem, proven by An. A. Muchnik, shows that allowing slightly super-logarithmic growth of the Kolmogorov complexity is enough to construct a sequence in $\PMR$.

\begin{theorem}[Muchnik et al.~\cite{MuchnikSU1998}]\label{theorem15}
For every computable order~$h$ there is a sequence $\seqa \in \mathbf{\PMR}$ such that, for all $n \in \mathbb{N}$, $$\Cc{\seqa \uh{n}}{n} \leq h(n)\log(n) + \ord{1}. $$
\end{theorem}

\begin{proof}
The proof is almost identical to the proof of Theorem~\ref{thm:facile-tmr}. The only difference is that we insert \emph{all} partial computable martingales one by one, and diagonalize against their weighted sum as before. It may happen however, that at some stage of the construction, one of the martingales becomes undefined. All we need to do then is to memorize this, and ignore this particular martingale from that point on. Call~$\seqa$ the sequence we obtain by this construction. We want to describe~$\seqa \uh{n}$. To do so, we need to specify~$n$, and, out of the~$k$ partial computable martingales that are inserted before stage~$n$, which ones have diverged, and at what stage, hence an information of $\ord{k \log n}$ (giving the position where a particular martingale diverges costs~$\ord{\log n}$ bits, and there are~$k$ martingales. Since we can insert martingales as slowly as we like (following some computable order), the complexity of $\seqa \uh{n}$ given~$n$ can be taken to be smaller than~$h(n)\log n+O(1)$ (where~$h$ is a computable order, fixed before the construction of~$\seqa$). 
\qed
\end{proof}

\subsection{The class $\PPR$}

In the case of total strategies, allowing permutation gives no real additional power, as $\TMR=\TPR$. Very surprisingly, Muchnik showed that in the case of partial computable strategies, permutation strategies are a real improvement over monotonic ones. To wit, the following theorem (quite a contrast to Theorem~\ref{theorem15}!).   

\begin{theorem}[Muchnik \cite{MuchnikSU1998}]\label{thm:muchnik-ppr}
If there is a computable order~$h$ such that for all~$n$ we have $\K(A \ut n) \leq n - h(n) - \ord{1}$, then $A \not\in \PPR$.
\end{theorem}

Note that the proof used by Muchnik in~\cite{MuchnikSU1998} works if we replace $\K$ by $\C$ in the above statement. 

\begin{theorem}\label{thm:ppr-low-complx}
For every computable order~$h$ there is a sequence $\seqa \in \PPR$, such that there are infinitely many $n$ where $\Cc{\seqa \uh{n}}{n} < h(n)$. 

Furthermore, if we have an infinite computable set $S \subseteq \mathbb{N}$, we can choose the infinitely many lengths $n$ such that they all are contained in $S$.
\end{theorem}

\begin{lemma}\label{lem:totalization}
Let~$d$ be a partial computable martingale. Let $\mathcal{C}$ be an effectively closed subset of 
~$\cs$. Suppose that~$d$ is total on every element of~$\mathcal{C}$. Then there exists a total computable martingale~$d'$ such that $\Succ(d) \cap \mathcal{C} = \Succ(d') \cap \mathcal{C}$. 
\end{lemma}

\begin{proof}
The idea of the proof is simple: the martingale~$d'$ will try to mimic~$d$ while enumerating the complement~$\mathcal{U}$ of~$\mathcal{C}$. If at some stage a cylinder~$[w]$ is covered by~$\mathcal{U}$, then~$d$ will be passive (i.e.\ defined but constant) on the sequences extending~$w$. As we do not care about the behavior of~$d'$ on~$\mathcal{U}$ (as long as it is defined), this will be enough to get the conclusion.\\

Let $d, \mathcal{C}$ be as above. We build the martingale~$d'$ on words by induction. Define $d'(\emptyword)=d(\emptyword)$ (here we assume without loss of generality that $d(\emptyword)$ is defined, otherwise there is nothing to prove). During the construction, some words will be marked as inactive, on which the martingale will be passive; initially, there is no inactive word. On active words~$w$, we will have~$d(w)=d'(w)$.\\

Suppose for the sake of the induction that $d'(w)$ is already defined. If $w$ is marked as inactive, we mark $w0$ and $w1$ as inactive, and set $d(w0)=d(w1)=d(w)$. Otherwise, by the induction hypothesis, we have $d(w)=d'(w)$. We then run in parallel the computation of~$d(w0)$ and~$d(w1)$, and enumerate the complement~$\mathcal{U}$ of~$\mathcal{C}$ until one of the two above events happens:

\begin{itemize}
\item[(a)] $d(w0)$ and $d(w1)$ become defined. Then set $d'(w0)=d(w0)$ and $d'(w1)=d(w1)$
\item[(b)] The cylinder $[w]$ gets covered by~$\mathcal{U}$. In that case, mark $w0$ and $w1$ as inactive and set $d'(w0)=d'(w1)=d'(w)$
\end{itemize} 

Note that one of these two events \emph{must} happen: indeed, if $d(w0)$ and $d(w1)$ are undefined (remember that by the definition of a martingale, Definition~\ref{def:martingale}, that they are either both defined or both undefined), then this means that~$d$ diverges on \emph{any} element of $[w0] \cup [w1]= [w]$. Hence, by assumption, $[w] \cap \mathcal{C} = \emptyset$, i.e.\ $[w] \subseteq \mathcal{U}$. It remains to verify that $\Succ(d) \cap \mathcal{C} = \Succ(d') \cap \mathcal{C}$. Let $\seqa \in \mathcal{C}$. Since $d$ is total on~$\seqa$ by assumption, during the construction of $d'$ on $\seqa$, we will always be in case~(a), hence we will have for all~$n$, $d(\seqa \uh{n}) = d'(\seqa \uh{n})$. The result follows immediately. \qed

\end{proof}

\begin{corollary}\label{cor:totalization2}
Let~$b=(d,\pi)$ be a partial computable permutation strategy (resp.\ injective strategy). Let~$\mathcal{C}$ be an effectively closed subset of~$\cs$. Suppose that~$b$ is total on every element of~$\mathcal{C}$. Then there exists a total computable permutation strategy (resp.\ injective strategy) $b'$ such that $\Succ(b) \cap \mathcal{C} = \Succ(b') \cap \mathcal{C}$. 
\end{corollary}

\begin{proof}
This follows from the fact that the image or pre-image of an effectively closed set under a computable permutation of the bits is itself a closed set: take $b=(d,\pi)$ and $\mathcal{C}$ as above. Let $\bar{\pi}$ be the map induced on $\cs$ by~$\pi$, i.e. the map defined for all~$\seqa \in \cs$ by
$$
\bar{\pi}(\seqa)=\seqa(\pi(0))\seqa(\pi(1))\seqa(\pi(2))\ldots
$$ 
For any given sequence~$\seqa \in \mathcal{C}$, $b$ succeeds on~$\seqa$ if and only if $d$~succeeds on~$\bar{\pi}(A)$. As $\bar{\pi}(A) \in \bar{\pi}(\mathcal{C})$, and  $\bar{\pi}(\mathcal{C})$ is an effectively closed set, by~\label{lem:totalization}, there exists a total martingale~$d'$ such that $\Succ(d) \cap \bar{\pi}(\mathcal{C}) = \Succ(d') \cap \bar{\pi}(\mathcal{C})$. Thus, $d'$ succeeds on $\bar{\pi}(A)$, or equivalently, $b'=(d',\pi)$ succeeds on~$A$. Thus $b'$ is as desired.  \qed

\end{proof}


\begin{proof}[of Theorem~\ref{thm:ppr-low-complx}]
Again, this proof is a variant of the proof of Theorem~\ref{thm:facile-tmr}: we add strategies one by one, diagonalizing, at each stage, against a finite weighted sum of total monotonic strategies (i.e.\ martingales). Of course, not all strategies have this property, but we can reduce to this case using the techniques we presented above. Suppose that in the construction of our sequence~$\seqa$, we have already constructed an initial segment~$w_k$, and that up to this stage we played against a weighted sum of~$k$ total martingales
$$
D_k=\sum_{i=1}^k \alpha_i \, d_i
$$
where the $d_i$ are total computable martingales, ensuring that $D(u) < 2$ for all prefix~$u$ of~$w$. Suppose we want to introduce a new strategy~$b=(d,\pi)$. There are three cases:\\

Case 0: the new strategy is not valid, i.e.\ $\pi$ is not a permutation. In this case, we just add one bit of extra information to record this, and ignore~$b$ from now on, i.e.\ we set~$w_{k+1}=w_k$, $d_{k+1}=0$ (the zero martingale), and $D_{k+1}=D_k+d_{k+1}=D_k$.\\

Case 1: the strategy~$b$ is indeed a partial computable permutation strategy, and there exists an extension~$w'$ of~$w$ such that $D_k(u) < 2$ for all prefixes~$u$ of $w'$, and $b$ diverges on~$w'$. In this case, we simply take $w'$ as our new prefix of~$\seqa$, as it both diagonalizes against~$D$, and defeats~$b$ (since~$b$ diverges on~$w'$, it will not win against \emph{any} possible extension of~$w'$). We can thus ignore~$b$ from that point on, so we set $w_{k+1}=w'$, $d_{k+1}=0$ and $D_{k+1}=D_k+d_{k+1}=D_k$.\\

Case 2: if we are not in one of the two previous cases, this means that our strategy~$b=(d,\pi)$ is a partial computable permutation strategy, and that~$b$ is total on the whole $\mathrm{\Pi}^0_1$ class 
$$
\mathcal{C}_k=[w_k] \cap \{X \in \cs \mid \forall n\, D_k(X \uh{n}) < 2\}
$$
Thus, by Lemma~\ref{cor:totalization2}, there exists a total computable permutation strategy~$b'$ such that $\Succ(b) \cap \mathcal{C}_k = \Succ(b') \cap \mathcal{C}_k$. And by Theorem~\ref{thm:permutation-monotonization}, there exists a total computable martingale~$d''$ such that~$\Succ(b') \subseteq \Succ(d'')$. Thus, we can replace~$b$ by $d''$, and defeating $d''$ will be enough to defeat~$b$ as long as the sequence we construct is in~$\mathcal{C}_k$. We thus set $d_{k+1}=d''$, $w_{k+1}=w_k$ and 
$$
D_{k+1}=\sum_{i=1}^{k+1} \alpha_i \, d_i
$$
where $\alpha_{k+1}$ is sufficiently small to have $D_{k+1}(w_{k+1}) < 2$.\\

Once we have added a new monotonic martingale, we (as usual) computably find an extension $w''$ of $w_{k+1}$, ensuring that $D_{k+1}(u) < 2$ for all prefix~$u$ of $w''$, taking $w''$ long enough to have $\Cc{w''}{|w''|} \leq h(|w''|)$. We then set $w_{k+1}=w''$, then add a $k+2$-th strategy and so on.\\

Note that since~$w''$ can be chosen arbitrarily large, if we have fixed a computable susbet $S$ of $\N$, we can also ensure that $|w''|$ belong to~$S$ if we like. 


It is clear that the infinite sequence~$A$ constructed via this process satisfies $$\Cc{\seqa \uh{n}}{n} \leq h(n)$$ for infinitely many~$n$ (and, since Case 2 happens infinitely often, if we fix a given computable set~$S$, we can ensure that infinitely many of such~$n$ belong to~$S$). To see that it belongs to~$\PPR$, we notice that since for all~$k$, $D_{k+1} \geq D_k$ and $w_k \prefx w_{k+1}$, we have $\mathcal{C}_{k+1} \subseteq \mathcal{C}_k$ and thus $A \in \bigcap_k \mathcal{C}_k$. Now, given a total computable strategy~$b=(d,\pi)$, let~$k$ be the stage where~$b$ was considered, and replaced by the martingale $d_k$. Since by construction of~$\seqa$, $d_{k+1}$ does not win against~$\seqa$ and by definition of~$d_k$, $\Succ(b) \cap \mathcal{C}_k \subseteq \Succ(d_k) \cap \mathcal{C}_k$, it follows that $\seqa \notin \Succ(b)$. \qed

\end{proof}

Now that we have assembled all our tools, we can easily prove the desired results.

\begin{theorem}
The following statements hold.
\begin{enumerate}
\item $\PPR \not\subseteq \TIR$\label{conclusion1}
\item $\TIR \not\subseteq \PMR$\label{conclusion2}
\item $\PMR \not\subseteq \PPR$
\end{enumerate}
From these results it easily follows that in Figure~\ref{figure:inclusionstructure} no inclusion holds except those indicated and those implied by transitivity.
\end{theorem}
\begin{proof}
\begin{enumerate}
\item Choose a computable sequence $\{n_k\}_k$ fulfilling the requirements of Theorem \ref{thm:tir1} such that $\C(k) \leq \log\log n_k$ for all $k$. The members of this set then form a computable set $S$. Use Theorem \ref{thm:ppr-low-complx} to construct a sequence $A \in \PPR$ such that $\C(A\ut n \mid n) < \log\log n$ at infinitely many places in $S$. We then have for infinitely many $k$
\[\C(A\ut n_k \mid k) \leq \C(A\ut n_k) \leq \C(A\ut n_k \mid n_k) + 2\log\log n_k \leq 3\log\log n_k,\]
so $A$ cannot be in $\TIR$ according to Theorem  \ref{thm:tir1}.
\item Follows immediately from Theorems \ref{thm:tir2} and \ref{theorem-merkle}.
\item Follows immediately from Theorems \ref{theorem15} and \ref{thm:muchnik-ppr}.
\qed
\end{enumerate}
\end{proof}

\begin{figure}[h]
\begin{center}
\framebox{%
\begin{tabular}{r|ccccc}
 &  monotonic && permutation && injection\\
 \hline

 total\, & \TMR & $=$ & \TPR &  $\supsetneq$ & \TIR\\
          &  \rotatebox{90}{$\subsetneq$}& & \rotatebox{90}{$\subsetneq$} & & \rotatebox{90}{$\subsetneq$}\\
 partial\, & \PMR & $\supsetneq$ & \PPR & $\supsetneq$ & \PIR\\
\end{tabular}
}
\end{center}
  \caption{Assembled class inclusion results \label{figure:inclusionstructure}}
\end{figure}

\newpage
\bibliography{Separations_of_non_monotonic_randomness_notions.Frame}
\bibliographystyle{plain}

\end{document}